\documentclass[aps,pra,twocolumn,groupedaddress,showpacs]{revtex4}

\usepackage[dvips]{graphicx}
\usepackage{epsfig}
\usepackage{amssymb}
\usepackage{amsmath}
\usepackage{braket}
\usepackage{color}

\begin{document}

% Use the \preprint command to place your local institutional report
% number in the upper righthand corner of the title page in preprint mode.
% Multiple \preprint commands are allowed.
% Use the 'preprintnumbers' class option to override journal defaults
% to display numbers if necessary
%\preprint{}

%Title of paper
%\title{Accelerating, decelerating and guiding fast beams of Rydberg atoms above a segmented transmission line}
\title{Transmission-line decelerators for atoms in high Rydberg states}

% repeat the \author .. \affiliation  etc. as needed
% \email, \thanks, \homepage, \altaffiliation all apply to the current
% author. Explanatory text should go in the []'s, actual e-mail
% address or url should go in the {}'s for \email and \homepage.
% Please use the appropriate macro for each type of information

% \affiliation command applies to all authors since the last
% \affiliation command. The \affiliation command should follow the
% other information
% \affiliation can be followed by \email, \homepage, \thanks as well.
\author{P. Lancuba and S. D. Hogan}
%\email[]{Your e-mail address}
%\homepage[]{Your web page}
%\thanks{}
%\altaffiliation{}

\affiliation{Department of Physics and Astronomy, University College London, Gower Street, London WC1E 6BT, U.K.}

%Collaboration name if desired (requires use of superscript address
%option in \documentclass). \noaffiliation is required (may also be
%used with the \author command).
%\collaboration can be followed by \email, \homepage, \thanks as well.
%\collaboration{}
%\noaffiliation

\date{\today}

\begin{abstract}
Beams of helium atoms in Rydberg states with principal quantum number $n=52$, and traveling with an initial speed of 1950~m/s, have been accelerated, decelerated and guided while confined in moving electric traps generated above a curved, surface-based electrical transmission line with a segmented center conductor. Experiments have been performed with atoms guided at constant speed, and with accelerations exceeding $10^7$~m/s$^2$. In each case the manipulated atoms were detected by spatially resolved, pulsed electric field ionization. The effects of tangential and centripetal accelerations on the effective trapping potentials experienced by the atoms in the decelerator have been studied, with the resulting observations highlighting contributions from the density of excited Rydberg atoms to the acceleration, deceleration and guiding efficiencies in the experiments.
\end{abstract}

% insert suggested PACS numbers in braces on next line
% insert suggested keywords - APS authors don't need to do this
%\keywords{}

%\maketitle must follow title, authors, abstract, \pacs, and \keywords
\maketitle

\section{Introduction}

Chip-based devices for controlling the translational motion of, and trapping, gas-phase samples of atoms~\cite{reichel11a}, molecules~\cite{meek08a} and ions~\cite{chiaverini05a} offer robust and scalable routes to the generation of complex trap structures, and opportunities for integration with other surface-based components, including for example, co-planar microwave waveguides or resonators~\cite{rabl06a,bohi10a,hogan12a,carter13a}. The recent development of chip-based architectures for manipulating atoms and molecules in high Rydberg states~\cite{hogan12a,hogan12b,allmendinger13a,allmendinger14a} has opened up exciting opportunities for (1) studies of the decay processes and interactions of Rydberg states of heavy molecules traveling in beams with kinetic energies that were too large for efficient manipulation using previous decelerator designs~\cite{hogan08a,seiler11a}; (2) investigations of the interactions between Rydberg atoms or molecules and surfaces~\cite{hill00a,lloyd05a}; (3) applications of Rydberg atoms in hybrid approaches to quantum information processing, in which the transport and confinement of samples above superconducting microwave circuits is required~\cite{hogan12a,thiele14a}; and (4) the manipulation of beams of positronium and antihydrogen atoms in long-lived high Rydberg states for gravitational and spectroscopic studies~\cite{kellerbauer08a,cassidy14a}. 

\begin{figure*}
\begin{center}
\includegraphics[width = 0.9\textwidth, angle = 0, clip=]{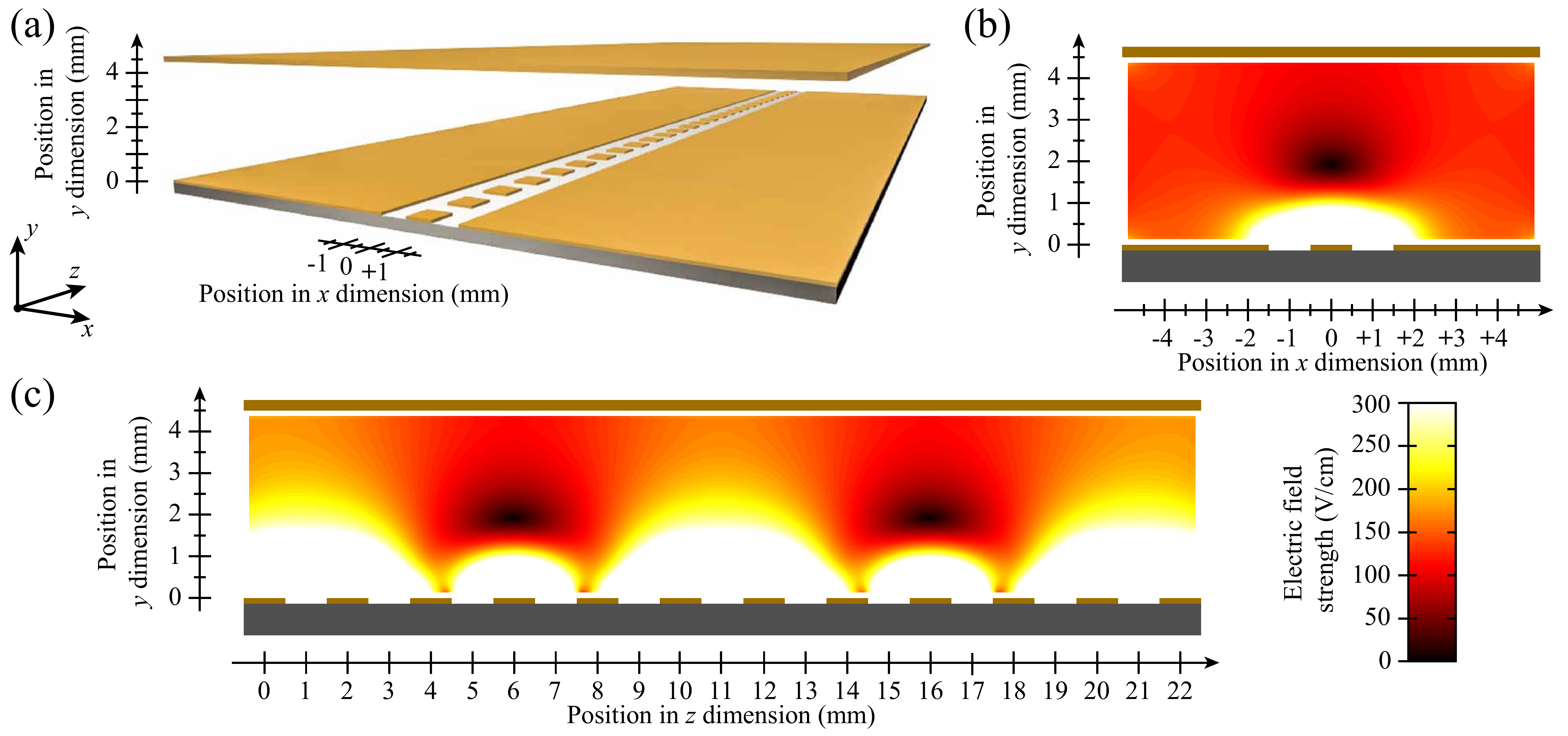}
\caption{(Color online) (a) Schematic diagram of the transmission-line decelerator. Electric field distributions in (b) the $xy$-plane, and (c) the $zy$-plane, for $V_{0}=+120$~V, and $V_{\mathrm{u}}=-V_{0}/2$. The color scale in the lower right of the figure is common to (b) and (c).}
\label{fig1}
\end{center}
\end{figure*}

The chip-based Rydberg-Stark decelerator described here, and referred to as a \emph{transmission-line decelerator}, relies upon the continuous motion of a periodic array of electric field minima in the void between a two-dimensional (2D) surface-based arrangement of metallic electrodes, and a parallel plane metal plate [see Fig.~\ref{fig1}(a)]. This decelerator differs from previously developed chip-based decelerators for polar ground-state molecules~\cite{meek08a,meek09a,marx13a} and for Rydberg atoms and molecules~\cite{hogan12b,allmendinger13a,allmendinger14a}, and other Stark decelerators that employ continuously moving electric traps~\cite{osterwalder10a,quinteroperez13a,meerakker12a}, in that it is based upon the geometry of a surface-based transmission-line guide~\cite{lancuba13a}. Such electrostatic guides have been used to guide and deflect fast beams of helium atoms while providing two-dimensional transverse confinement. However, they do not provide longitudinal confinement and cannot be used to accelerate or decelerate an atomic or molecular beam. To transform one of these guides into a device suitable for acceleration, or deceleration, of samples traveling in pulsed supersonic beams, the center-conductor has been divided into discrete, individually addressable segments. This geometry permits the generation of three-dimensional electric traps that provide strong confinement of atoms or molecules in all three spatial dimensions throughout the decelerator. It is compatible with the introduction of curvatures in the plane of the 2D electrode array without causing excessive distortion of the fields, and with direct coupling to transmission-line guides and beam-splitters. It also provides a well defined electromagnetic environment in which trapped Rydberg atoms or molecules can be transported and manipulated. This is essential for investigating, and controlling, effects of blackbody radiation. 

As in other Rydberg-Stark decelerators, the transmission-line decelerator described here relies upon the forces exerted by inhomogeneous electric fields on atoms or molecules in Rydberg states with linear Stark energy shifts, i.e., states with permanent effective electric dipole moments~\cite{wing80a,breeden81a,yamakita04a,vliegen04a,hogan09a}. For each value of the principal quantum number $n$, the magnitude of the maximal associated electric dipole moment is $\mu_{\mathrm{max}}\simeq(3/2)n^2e\,a_0$, where $e$ is the charge of the electron, and $a_0$ is the Bohr radius corrected for the reduced mass of the system~\cite{gallagher94a}. For states with principal quantum numbers close to $n=50$ these dipole moments exceed 9\,000~D. In the transmission-line decelerator, the periodic array of electric field minima generated above the segmented center-conductor act as traps for samples in low-field-seeking Rydberg-Stark states, with positive Stark energy shifts. Upon the application of appropriate time-dependent electric potentials to the decelerator electrodes these traps can be conveyed through the device, allowing the samples confined within them to be accelerated, decelerated and guided.

In the following, the design and operation principles of the transmission-line decelerator are first presented. An overview of the experimental apparatus in which the decelerator was installed to perform the experiments reported here is then provided. Thereafter, the results of experiments in which beams of helium (He) atoms in Rydberg-Stark states with $n=52$, and traveling with an initial speed of 1950~m/s, were guided at constant speed through the decelerator are presented. Finally, the results of experiments involving the acceleration and deceleration of these atoms to final speeds ranging from 2700~m/s to 1350~m/s are discussed.\\

\section{Decelerator design}\label{sec:dec}

A schematic diagram of the transmission-line decelerator is displayed in Fig.~\ref{fig1}(a). The design, based upon that of an electrostatic transmission-line guide~\cite{lancuba13a}, includes a 2D surface-based array of metallic electrodes, and a parallel plane metal plate. The 2D electrode array, located 4.5~mm below the upper plate, is composed of a segmented center-conductor surrounded on either side by two extended ground planes. The insulating gap between the center-conductor and each ground plane is 1~mm wide. The segments of the center-conductor are square, with sides of length 1~mm, and a center-to-center spacing $d_{\textrm{cc}}=2$~mm. The electrode array used in the experiments described here was fabricated by chemical etching. Electrical contact was made to the individual decelerator segments via small holes drilled through each to an array of miniature sockets located beneath the substrate. The center conductor of this device contained 45 segments, and the array had a total length of 90~mm in the $z$-dimension, and a width of 70~mm in the $x$-dimension. 

With equal, non-zero electric potentials applied to one single segment of the center-conductor and the upper plate, while all other electrodes including the ground planes are set to $0$~V, a single stationary three-dimensional electric field minimum is generated above this active segment. The resulting electric field distribution is analogous to that generated in two-dimensions in an electrostatic transmission-line guide~\cite{lancuba13a}, and forms the basic field distribution around which the decelerator is designed. By applying an appropriate set of potentials to the complete 2D array of decelerator segments and upper plate, a periodic array of similar electric field minima can be generated within the device. The spatial periodicity of this array of electric field minima is dependent upon the number of electric potentials, and therefore decelerator segments, contributing to each. Exploiting larger numbers of segments leads to increasingly harmonic traps with reduced variations in the trap-width with position along the decelerator axis. In the decelerator described here a set of five electric potentials are employed to achieve an appropriate compromise between the total number of unique potentials that must be generated, and any undesirable position dependence of the trap-widths. These five potentials are applied to consecutive decelerator segments and repeated periodically along the center conductor. The magnitude of the potential applied to each electrode, labeled with an index $i=1,2,\dots$, can be expressed as
\begin{equation}
V_{i}=V_{0}\,\cos[- (i-1)\,\phi],\label{eq:timeindeppot}
\end{equation}
where $V_{0}$ is the amplitude of the potential applied to electrode~1, and $\phi=2\,\pi/5$ is the phase-shift from one decelerator segment to the next. With a dc potential, $V_{\textrm{u}}$, applied to the upper plate, such that $V_{\textrm{u}}=-V_{0}/2$, a set of electric field minima is generated with a spatial periodicity of $5\,d_{\textrm{cc}}$. Electric field distributions calculated using this configuration of potentials, with $V_{0}=+120$~V, and a phase shift of $\pi/5$, to position the electric field minima above electrodes $i=4,9,\,\dots$, are displayed in Fig.~\ref{fig1}(b), and Fig.~\ref{fig1}(c). In these figures, each electric field minimum represents a trap for atoms in low-field-seeking Rydberg-Stark states. Close to each trap minimum the electric field gradient is $\sim650$~V/cm$^{2}$, and the lowest saddle-point of each trap occurs in the $y$-dimension at a field strength of $\sim110$~V/cm. The maximum usable value of $V_0$ is set by the onset of electric field ionization of the trapped Rydberg atoms and is therefore $n$-dependent. For the Rydberg-Stark states with $n=52$ used in the experiments reported here, complete diabatic electric field ionization occurs at $\sim140$~V/cm~\cite{gallagher94a}. By operating the decelerator with $V_{0}=+120$~V the fields experienced by the atoms as they travel through the decelerator are maintained below this value, limiting particle loss by direct field ionization.

\begin{figure}
\begin{center}
\includegraphics[width = 6.5cm, angle = 0, clip=]{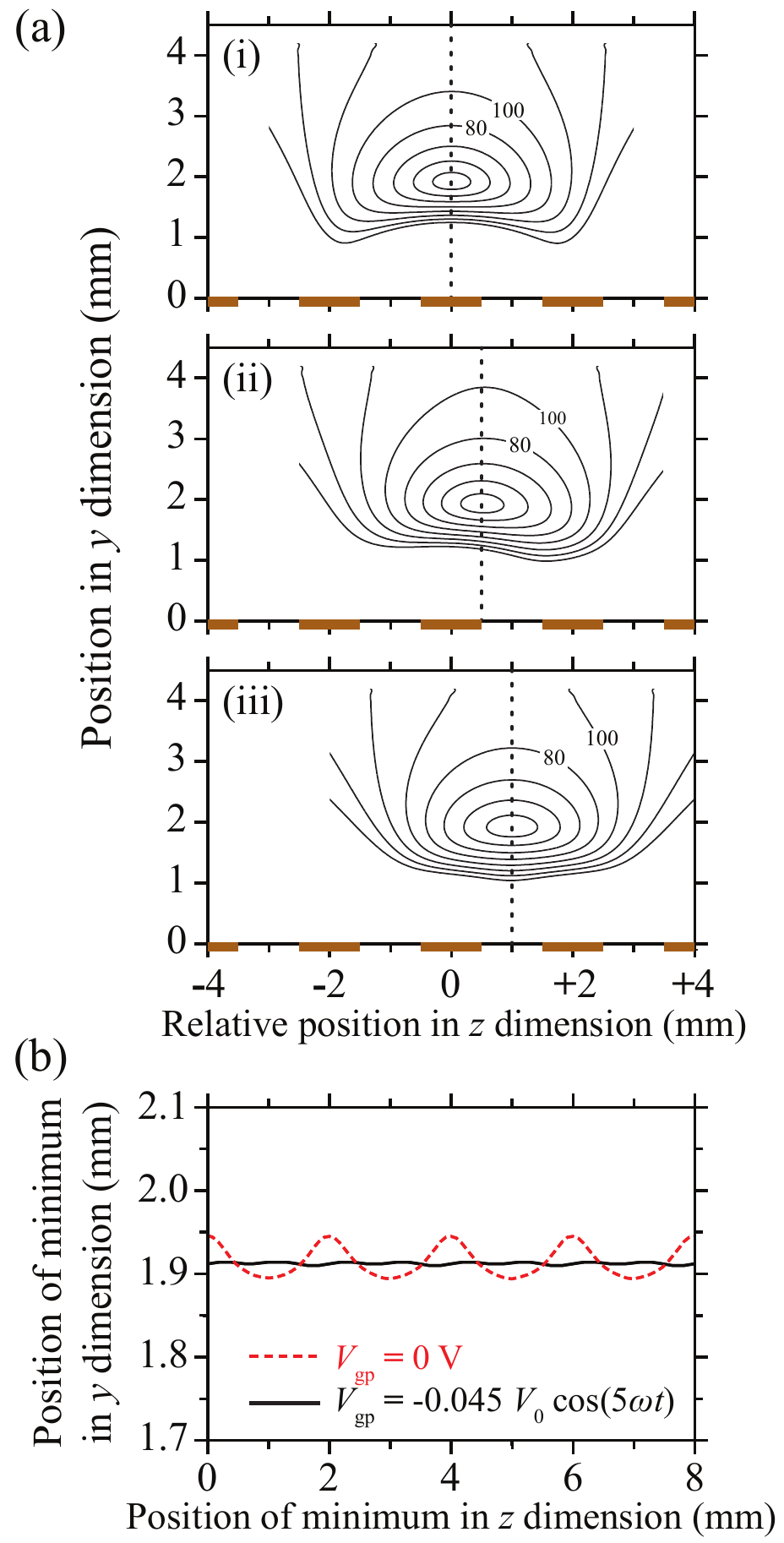}
\caption{(Color online) (a) Electric field distributions in the $zy$-plane for traps located (i) above a decelerator segment, (ii) one quarter of the way between two segments, and (iii) half-way between two segments. In each case contours of equal electric field strength are displayed in increments of 20~V/cm, from 20~V/cm to 160~V/cm, and the positions of the decelerator segments are indicated by the thick bars on the horizontal axis. (b) The dependence of the position of the electric field minimum in the $y$-dimension upon its location along the decelerator axis when a potential $V_{\textrm{gp}}=0$~V is applied to the ground planes (dashed red curve), and when the potential $V_{\textrm{gp}}=-0.045V_0\cos(5\omega t)$ is applied (continuous black curve).}
\label{fig2}
\end{center}
\end{figure}

To convey samples of trapped Rydberg atoms through the decelerator, it is necessary to vary the positions of the electric traps in time. This is achieved by introducing a time-dependence to the potential in Eq.~(\ref{eq:timeindeppot}). To realize a continuous motion of the array of traps in the positive $z$-dimension, the potential applied to the decelerator electrodes is chosen to oscillate at an angular frequency $\omega$, such that
\begin{equation}
V_{i}(t)=V_{0}\,\cos[\omega\,t- (i-1)\,\phi].\label{eq:timedeppot}
\end{equation}
The speed with which the traps then move, $v_{\textrm{trap}}$, is directly proportional to $\omega$, $d_{\mathrm{cc}}$, and the spatial periodicity of the array, i.e., $v_{\textrm{trap}}=5d_{\textrm{cc}}\omega/(2\,\pi)$.

As the traps move through the decelerator in the positive $z$-dimension, their proportions change slightly as they travel from locations directly above a decelerator segment to those between two segments. This can be seen in Fig.~\ref{fig2}(a), where electric field distributions in the $zy$-plane containing an electric field minimum are displayed at instants in time when a trap is located (i) directly above a decelerator segment, (ii) one quarter of the way between two segments, and (iii) half-way between two segments. When the trap is located directly above a decelerator segment it is deeper than when located half-way between two segments. The resulting breathing motion leads to an oscillation in the full-width-at-half-maximum of the traps of $\sim10\%$, and consequently a small amount of heating of the trapped atoms. However, if cold samples are confined close to a trap minimum heating effects arising from this breathing motion can be minimized. 

When a potential $V_{\textrm{gp}}=0$~V is applied to the ground planes of the decelerator, with $V_{0}=+120$~V, the positions of the trap minima oscillate in the $y$-dimension as they travel through the decelerator. This oscillatory motion has a peak-to-peak amplitude of $\sim50~\mu$m as can be seen in Fig.~\ref{fig2}(b). If not compensated, these oscillations give rise to a non-zero center-of-mass motion of the ensemble of trapped atoms in this dimension. This motion of the electric field minimum can be stabilized through the application of a time-dependent potential, $V_{\textrm{gp}}$, to the ground planes. For the decelerator described here, the optimal ground-plane potential was found to be one that oscillates in-phase with $V_1$ at a frequency $\omega_{\textrm{gp}}=5\,\omega$, and with an amplitude of $-0.045\,V_{0}$ such that,
\begin{equation}
V_{\textrm{gp}}(t)=-0.045\,V_{0}\,\cos(5\,\omega t).\label{eq:gppot}
\end{equation}
The higher oscillation frequency of this potential, than those applied to the segments of the decelerator center-conductor, arises from the requirement that an equal correction to the position of the traps in the $y$-dimension is applied every time a trap is located above (between etc \dots) a decelerator segment. Therefore the period of the oscillation of $V_{\mathrm{gp}}$ must equal the time taken for the traps to travel from one segment to the next, a distance $d_{\mathrm{cc}}$. With this potential applied to the ground planes, the oscillation of the position of the trap minima in the $y$-dimension is reduced to an amplitude of less than $10~\mu$m as indicated by the continuous black curve in Fig.~\ref{fig2}(b).

When operated at a constant frequency the transmission-line decelerator can be used to convey trapped atoms or molecules at constant speed. To accelerate or decelerate the samples, it is necessary to change the frequency of the oscillating electric potentials in time, introducing a frequency chirp. For a selected initial speed, $v_{\textrm{trap}}(0)$, and an acceleration, $a_{\textrm{trap}}$, the required time-dependent oscillation frequency can be expressed as
\begin{equation}
\omega(t)=\omega(0)+\frac{\pi a_{\textrm{trap}}}{5\,d_{\textrm{cc}}}\,t,\label{eq:timedepfreq}
\end{equation}
where $\omega(0)=2\pi v_{\textrm{trap}}(0)/(5\,d_{\textrm{cc}})$ is the initial angular frequency.

\begin{figure}
\begin{center}
\includegraphics[width = 0.42\textwidth, angle = 0, clip=]{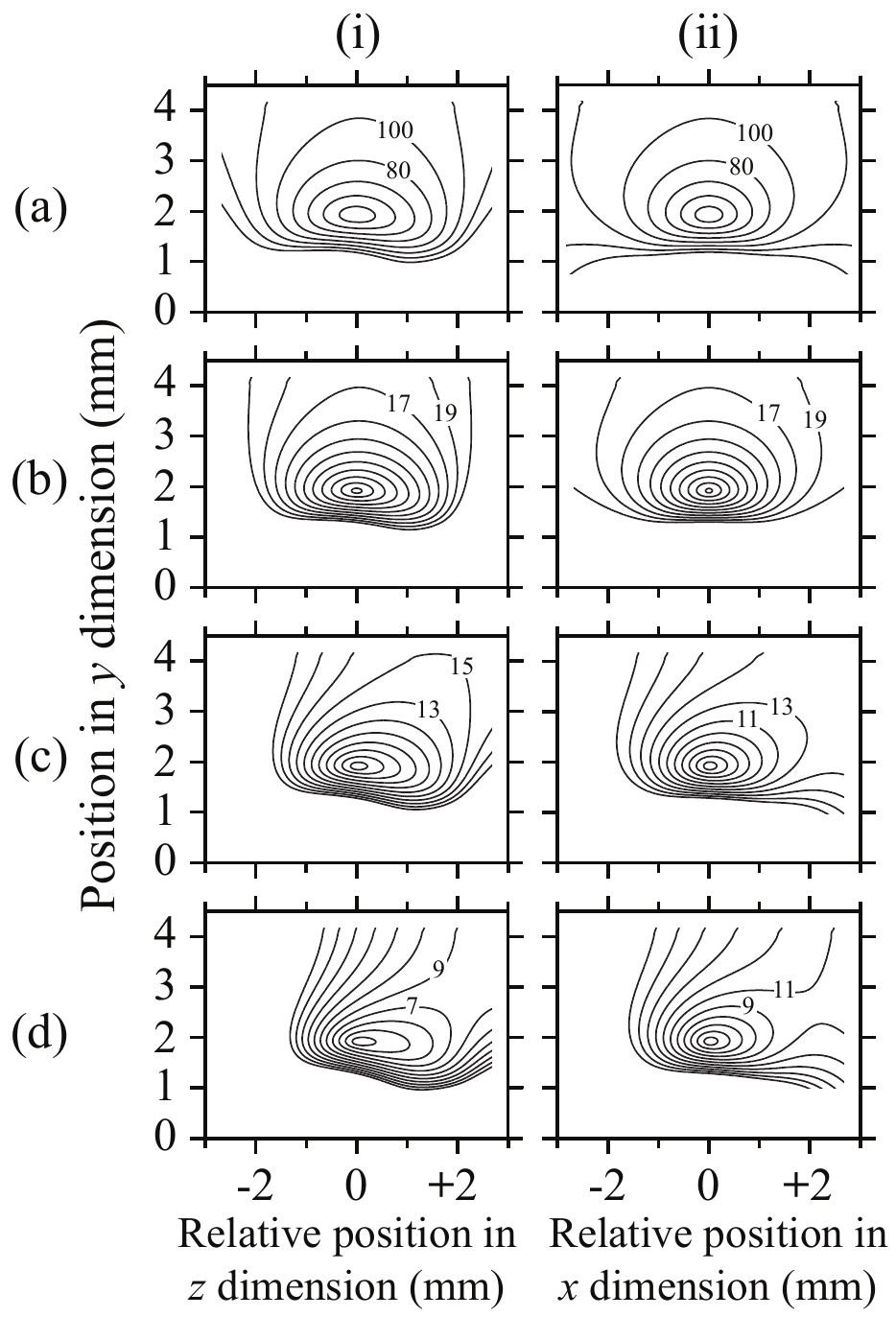}
\caption{(a) Electric field distributions in (i) the $zy$-plane, and (ii) the $xy$-plane, containing an electric field minimum located one quarter of the way between two decelerator segments. Contours of equal electric field strength are displayed in increments of 20~V/cm beginning at 20~V/cm. (b-d) Potential energy distributions in (i) the $zy$-plane, and (ii) the $xy$-plane, experienced by a He atom in the $\ket{n,k}=\ket{52,35}$ Rydberg-Stark state.  In (c-i) and (d-i), a pseudo potential is included to account for tangential accelerations of $-0.5\times10^7$~m/s$^2$ and $-1.15\times10^7$~m/s$^2$, respectively. In (c-ii) and (d-ii), the effects of centripetal accelerations of $-0.47\times10^7$~m/s$^2$ and $-0.71\times10^7$~m/s$^2$ are included. In the curved decelerator described here, these centripetal accelerations correspond to tangential velocities of $1950$~m/s and $2400$~m/s, respectively. In (b-d) contour lines of equal potential energy are displayed in terms of a temperature, i.e., $E/k_{\textrm{B}}$, in increments of 2~K beginning at 1~K.}
\label{fig3}
\end{center}
\end{figure}

The introduction of a frequency chirp to accelerate, or decelerate, trapped samples leads to a reduction in the effective depth of the traps. This has been discussed previously in the context of chip-based decelerators for polar ground-state molecules~\cite{meek09b}, with the consequences most clearly seen in the moving frame-of-reference associated with an individual electric field minimum. In Fig.~\ref{fig3}(a-i), and Fig.~\ref{fig3}(a-ii), the electric field distributions associated with a single trap of the decelerator are displayed in the $zy$- and $xy$-planes containing the electric field minimum. These electric field distributions have been calculated for the case when the trap is located one quarter of the way between two decelerator segments, a position that is close to the most probable location of the traps as they travel through the decelerator, with $V_0=+120$~V.

For a He atom in a low-field-seeking Rydberg-Stark state with an electric dipole moment of 6939~D, i.e., the $\ket{n,k}=\ket{52,35}$ state ($k$ is the difference between the two parabolic quantum numbers $n_1$ and $n_2$~\cite{gallagher94a}) used in the experiments reported here, the corresponding potential energy distributions are displayed in terms of a temperature, i.e., $E/k_{\mathrm{B}}$, in Fig.~\ref{fig3}(b-i), and Fig.~\ref{fig3}(b-ii). For atoms in the $\ket{52,35}$ state the traps have a depth of $E/k_{\textrm{B}}\simeq18$~K ($\simeq1.55$~meV) when stationary or moving at constant speed. 

In the moving frame of reference associated with an electric field minimum in the decelerator, the effect of the acceleration, $\vec{a}_{\mathrm{trap}}$, of the traps can be accounted for through the introduction of a pseudo-potential
\begin{equation}
V_{\textrm{pseudo}}=m\,\vec{a}_{\textrm{trap}}\cdot\vec{s},
\end{equation}
where $m$ is the mass of an individual atom, and $\vec{s}$ is the displacement from the electric field minimum. In the experiments described here a curved decelerator was employed to reduce effects of collisions of the accelerated, or decelerated, atoms with the surrounding components of the pulsed supersonic beam~\cite{seiler11a}. The curvature was introduced in the plane of the 2D electrode array. The radius of curvature of $\sim0.81$~m that was selected resulted in the exit of the decelerator being displaced by 5~mm in the negative $x$-dimension with respect to the entrance. In this curved device, acceleration of the trapped particles therefore takes place in two directions. These are: (1) the acceleration tangential to the curvature of the decelerator, $a_{\mathrm{t}}$, that is exploited to modify the longitudinal motion of the atoms, and (2) the centripetal acceleration, $a_{\mathrm{c}}$, associated with the curvature of the decelerator. The effects of these two accelerations can be investigated by studying the shape of the moving traps in the $zy$- and $xy$-planes, respectively.

For a small section of the decelerator, as considered in Fig.~\ref{fig3}, the tangential acceleration acts along the decelerator axis, i.e., in the $z$-dimension. The effect of this acceleration can therefore be accounted for by adding the pseudo potential $V_{\textrm{pseudo}}^{\mathrm{\,t}}=m\,a_{\mathrm{t}}\,\Delta z$ to the potential energy distribution in Fig.~\ref{fig3}(b-i). For He atoms in the $\ket{n,k}=\ket{52,35}$ Rydberg-Stark state, accelerations on the order of $10^6$~m/s$^2$ begin to significantly distort the unperturbed potential. The effect of tangential accelerations of $a_{\mathrm{t}} = -0.5\times10^7$~m/s$^2$, and $a_{\mathrm{t}} = -1.15\times10^7$~m/s$^2$ can be seen in Fig.~\ref{fig3}(c-i), and Fig.~\ref{fig3}(d-i), respectively. In Fig.~\ref{fig3}(c-i) the negative tangential acceleration (deceleration) leads to a reduction in the effective depth of the trap in the forward direction which becomes more pronounced as the magnitude of the acceleration is increased. When the traps are decelerated with $a_{\mathrm{t}} = -1.15\times10^7$~m/s the saddle point of the trap in the forward direction is further reduced, from $\sim18$~K when stationary, to $\sim8$~K. For an acceleration of $a_{\mathrm{t}} = -2.2\times10^7$~m/s$^2$ the trap opens completely in the forward direction, setting the limit on the tangential acceleration at which the decelerator can be operated for atoms in this Rydberg-Stark state. 

\begin{figure}
\begin{center}
\includegraphics[width = 8.5cm, angle = 0, clip=]{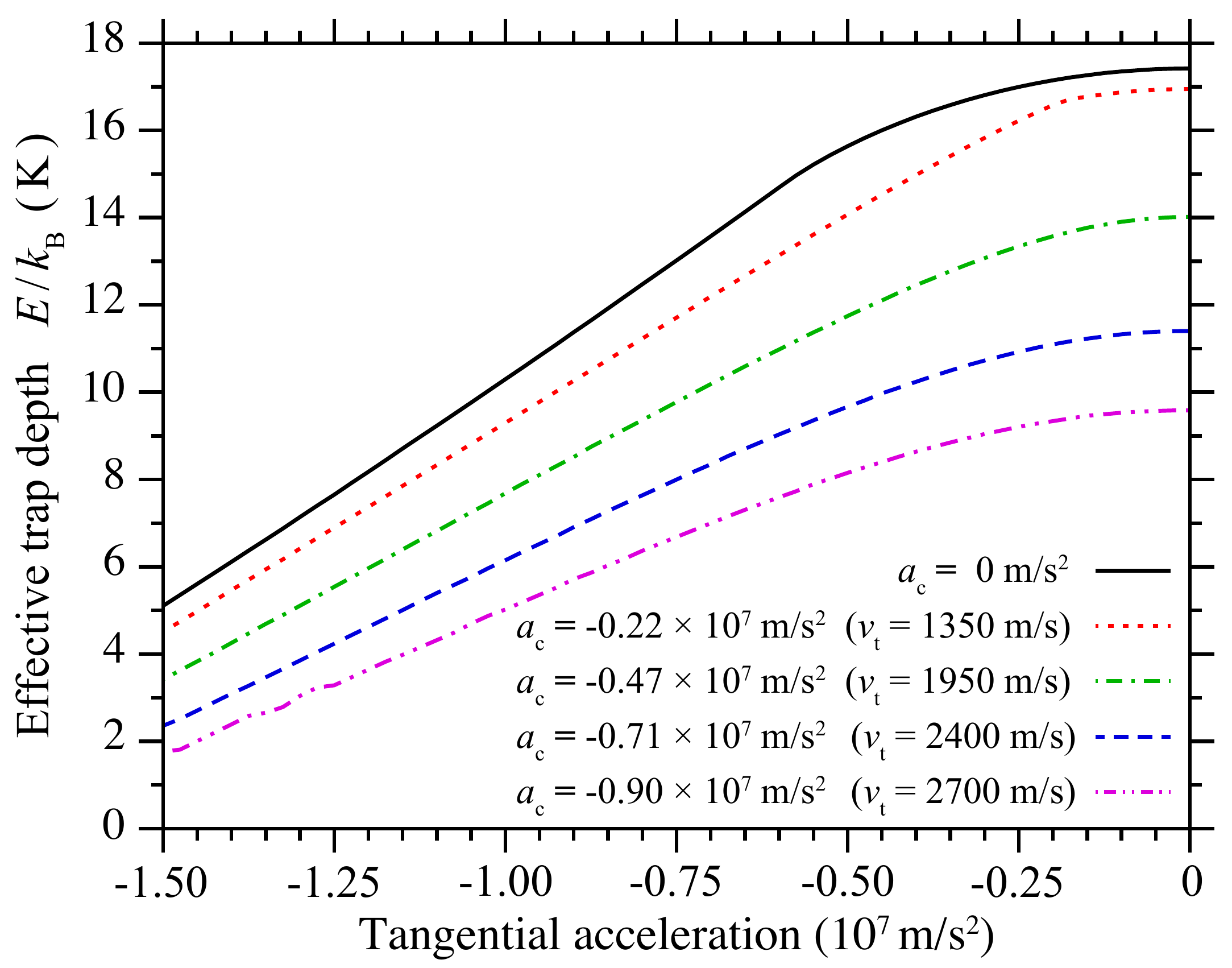}
\caption{(Color online) Effective depth of the moving decelerator traps for He atoms in the $\ket{n,k}=\ket{52,35}$ Rydberg-Stark state, when $V_{0}=+120$~V. Data for five different centripetal accelerations, $a_{\mathrm{c}}=0$~m/s$^2$ ($v_{\mathrm{t}}=0$~m/s -- continuous black curve), $a_{\mathrm{c}}=-0.22\times10^7$~m/s$^2$ ($v_{\mathrm{t}}=1350$~m/s -- dotted red curve), $a_{\mathrm{c}}=-0.47\times10^7$~m/s$^2$ ($v_{\mathrm{t}}=1950$~m/s -- dash-dotted green curve), $a_{\mathrm{c}}=-0.71\times10^7$~m/s$^2$ ($v_{\mathrm{t}}=2400$~m/s -- dashed blue curve), and  $a_{\mathrm{c}}=-0.90\times10^7$~m/s$^2$ ($v_{\mathrm{t}}=2700$~m/s -- dash--double-dotted magenta curve), are displayed.}
\label{fig4}
\end{center}
\end{figure}

The centripetal acceleration in a similar length of a curved decelerator acts in the direction perpendicular to the decelerator axis, i.e., in the $x$-dimension in Fig.~\ref{fig3}. In this case the effect of the acceleration can be accounted for by adding the pseudo potential $V_{\textrm{pseudo}}^{\mathrm{\,c}}=m\,a_{\mathrm{c}}\,\Delta x$, to the potential energy distribution in Fig.~\ref{fig3}(b-ii). For a radius of curvature of 0.81~m, centripetal accelerations on the order of $10^6$~m/s$^2$ occur for tangential velocities exceeding 1000~m/s. The effects of centripetal accelerations of $a_{\mathrm{c}} = -0.47\times10^7$~m/s$^2$, and $a_{\mathrm{c}} = -0.71\times10^7$~m/s$^2$, which correspond to tangential velocities of 1950~m/s and 2400~m/s, respectively, are displayed in Fig.~\ref{fig3}(c-ii), and Fig.~\ref{fig3}(d-ii). In the experiments reported here the initial longitudinal speed of the beams of He Rydberg atoms was 1950~m/s. Therefore even when the traps of the decelerator are moving at this constant initial speed the centripetal acceleration already plays an important role in setting the effective trap depth of $\sim14$~K. For the conditions under which the potential energy distributions in Fig.~\ref{fig3} were calculated, the maximal tangential velocity at which the decelerator can be operated before the traps open completely in the $x$-dimension is 4360~m/s. This tangential velocity corresponds to a centripetal acceleration of $a_{\mathrm{c}} = -2.35\times10^7$~m/s$^2$. An overview of the variation of the depth of the moving traps, for a range of tangential velocities and accelerations relevant to the experiments described, is presented in Fig.~\ref{fig4}. In this figure each effective trap depth was calculated using the complete three-dimensional potential energy distribution surrounding one electric field minimum within the decelerator.\\

\section{Experiment}\label{sec:exp}

\begin{figure}
\begin{center}
\includegraphics[width = 0.48\textwidth, angle = 0, clip=]{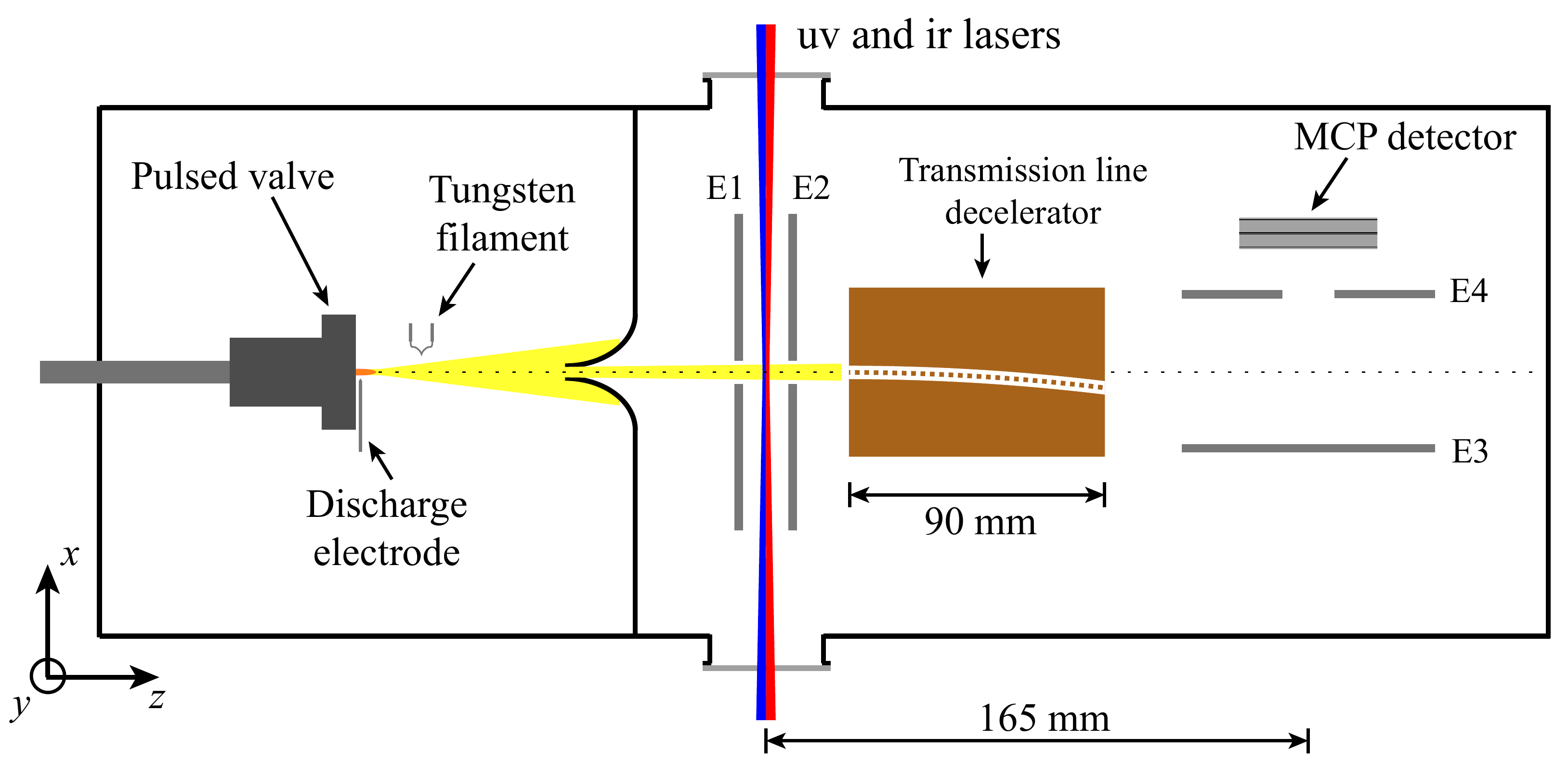}
\caption{(Color online) Schematic diagram of the experimental apparatus (not to scale). Photoexcitation of the metastable He atoms to selected Rydberg-Stark states is carried out between the electrodes labeled E1 and E2. After traveling through the decelerator the Rydberg atoms are detected by pulsed electric field ionization upon the application of a potential of $+3.5~$kV to the electrode labeled E3. The resulting He$^+$ ions are then accelerated through the aperture in the grounded electrode labeled E4 to an MCP detector.}
\label{fig5}
\end{center}
\end{figure}

A schematic overview of the apparatus used in the experiments reported here is presented in Fig.~\ref{fig5}. A pulsed supersonic beam of He atoms in the long-lived, metastable $1\mathrm{s}2\mathrm{s}\,^{3}\textrm{S}_{1}$ state, with an initial longitudinal speed of $1950$~m/s, was generated in an electric discharge at the exit of a pulsed valve \cite{halfmann00a}. After passing through a copper skimmer, the metastable atoms were photoexcited to Rydberg states with principal quantum number $n=52$, in a resonant two-photon excitation scheme via the 3p state using focused, co-propagating cw laser beams in the ultraviolet ($\lambda_{\textrm{uv}}\simeq389$~nm), and infrared ($\lambda_{\textrm{ir}}\simeq785$~nm) regions of the electromagnetic spectrum~\cite{lancuba13a}. The photoexcitation region in the apparatus was located between two parallel metallic plates labeled E1 and E2 in Fig.~\ref{fig5}. The experiments reported here required the excitation of only a small subensemble of the complete atomic beam in a precisely defined time window to selected Rydberg-Stark states. To achieve this a dc offset potential of $+0.30$~V was initially applied to E1, while E2 was set to $0$~V. This gave rise to a background electric field of $\sim0.23$~V/cm at the position where the laser beams crossed the atomic beam. The frequency of the infrared laser was then set to lie resonant with the transition to the low-field-seeking $\ket{n,k}=\ket{52,35}$ Rydberg-Stark state in a larger field of $0.61$~V/cm, so that atoms could be excited for selected periods of time by rapidly switching the potential on E1 from $+0.30$~V to $+0.80$~V, switching the electric field in the excitation region from its offset value to $0.61$~V/cm. In all of the experiments described here excitation was performed for a period of $5~\mu$s in each experimental cycle. The time associated with the middle of this excitation pulse is defined as the photoexcitation time, $t_{\mathrm{ex}}$. In the photoexcitation process the linear polarization of the two laser beams was chosen to permit the predominant excitation of hydrogenic Rydberg-Stark states with azimuthal quantum number $|m_{\ell}| = 2$. 

After photoexcitation, the excited Rydberg atoms passed through a small aperture in electrode E2 and entered the transmission-line decelerator within which they were accelerated, decelerated or guided. In the final section of the apparatus, beyond the decelerator, the atoms were detected by pulsed electric field ionization upon the application of a pulsed potential of $+3.5$~kV to E3. The He$^+$ ions produced following ionization of the excited atoms in the resulting electric field of $\sim550$~V/cm were then accelerated toward a microchannel plate (MCP) detector. The detection geometry employed permitted the position of the Rydberg atoms in the $x$-dimension, at the time of pulsed electric-field ionization, to be mapped onto the flight-time of the He$^+$ ions to the MCP detector, i.e., the ion signal corresponding to atoms that were located close to (far from) the MCP detector at the time of ionization arrived at early (late) times.

\begin{figure}
\begin{center}
\includegraphics[width = 0.42\textwidth, angle = 0, clip=]{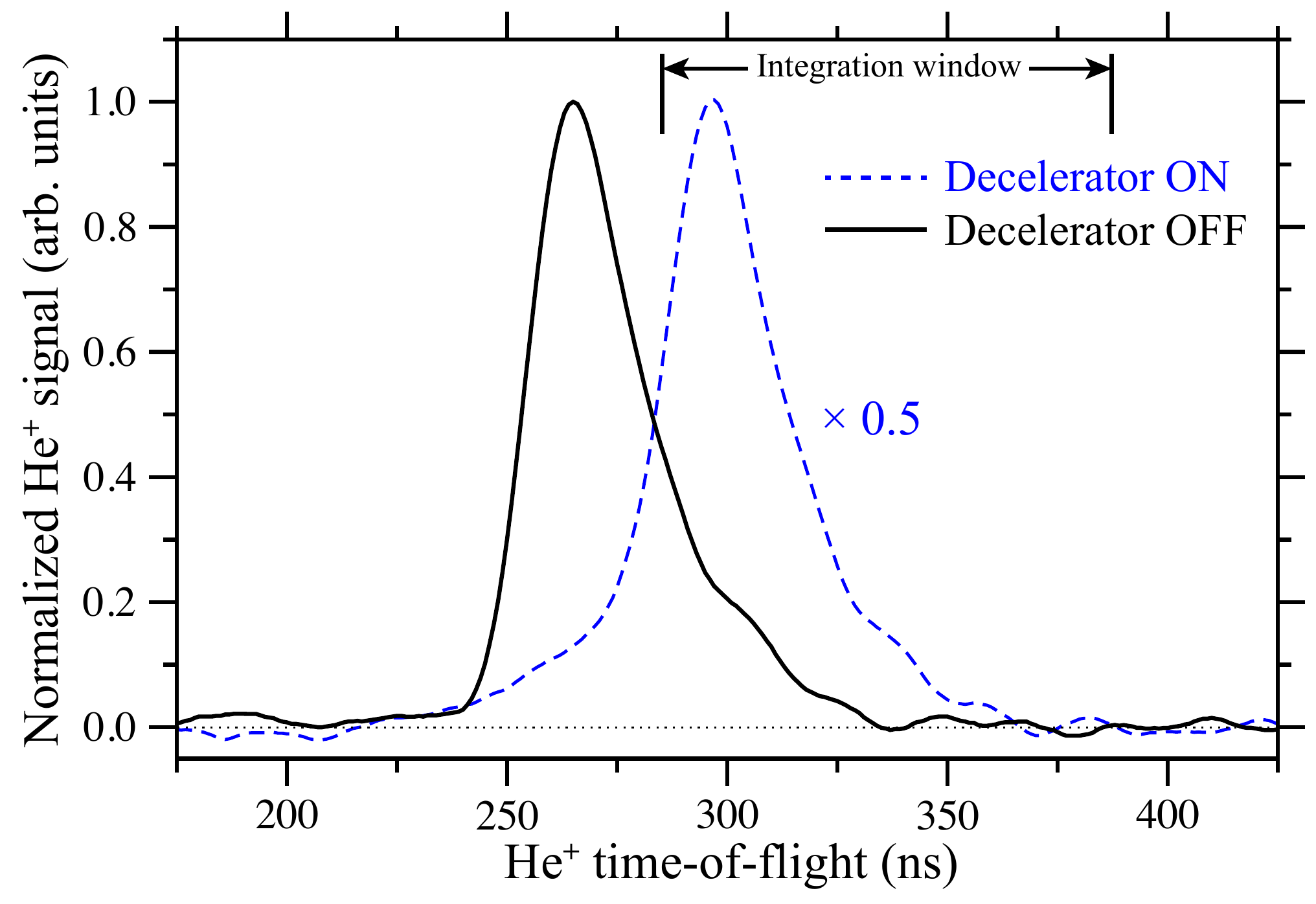}
\caption{(Color online) He$^+$ time-of-flight distributions recorded following pulsed electric field ionization of the He Rydberg atoms. When the decelerator is operated to guide atoms at a constant speed of 1950~m/s, the maximum of the time-of-flight distribution (dashed blue curve) arrives $\sim25$~ns later than that recorded with the decelerator inactive (continuous black curve). The time-of-flight distribution recorded with the decelerator active has been scaled by a factor of 0.5 (see text for details).}
\label{fig6}
\end{center}
\end{figure}

An example of the mapping of ionization position onto He$^+$ flight-time to the MCP is presented in Fig.~\ref{fig6}. When the oscillating decelerator potentials are off, but the non-zero dc potential $V_{\mathrm{u}}=-120/2$~V is applied to the upper plate electrode, the excited Rydberg atoms are ionized at a position in the detection region close to the MCP detector. The resulting He$^+$ time-of-flight distribution has a maximum at 265~ns. However, if the oscillating decelerator potentials are activated, and set to guide atoms at constant speed over the full length of the decelerator, the curvature of the device leads to a displacement of the Rydberg atoms into the negative $x$-dimension at the detection position. Because the atoms are then located further from the MCP detector when ionized the He$^+$ flight-time increases to 290~ns. Accelerated, decelerated and guided atoms can therefore be selectively detected by integrating, and recording, only the delayed component of the time-of-flight distribution encompassed by the vertical bars at the top of Fig.~\ref{fig6}. In the data displayed in this figure the signal recorded with the decelerator active is twice as large as that recorded with the oscillating decelerator potentials off because of the increased blackbody photoionization, and tunnel ionization rates in the residual dc electric field of 133~V/cm present in the latter case.\\

\begin{figure}
\begin{center}
\includegraphics[width = 0.48\textwidth, angle = 0, clip=]{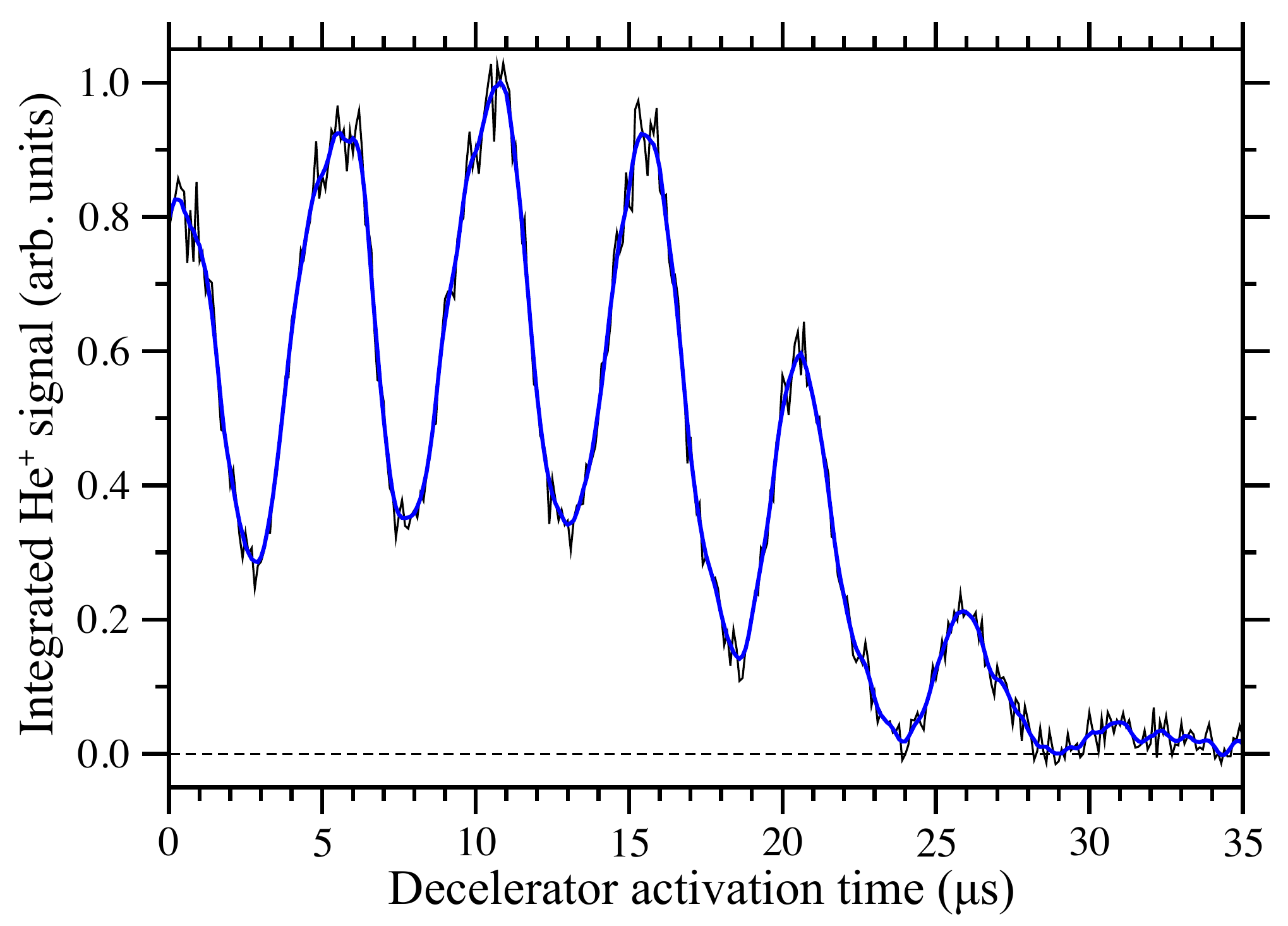}
\caption{(Color online) Dependence of the integrated He$^+$ signal on the activation time of the oscillating decelerator potentials, for a He atom flight-time from the photoexcitation region to the detection region of $82.5~\mu$s, when the decelerator is operated in a guiding mode at a constant speed of $1950~$m/s.}
\label{fig7}
\end{center}
\end{figure}

\section{Results}\label{sec:res}

To demonstrate and characterize the operation of the transmission-line decelerator, experiments have been performed in three modes of operation: (1) guiding at constant speed, (2) acceleration, and (3) deceleration. 

For each type of experiment it was necessary to ensure that the oscillating decelerator potentials were activated at the optimal time to ensure that the $5~\mu$s ($\sim10$~mm) long bunches of He Rydberg atoms prepared in the photoexcitation region were efficiently loaded into a single moving trap. This was achieved by measuring the integrated signal from atoms guided through the curved decelerator at constant speed, as the activation time was varied. This measurement was therefore performed after a fixed Rydberg atom flight-time from the photoexcitation region to the detection region, while recording only the signal in the delayed component of the He$^+$ time-of-flight distribution (see Fig.~\ref{fig6}), with the results displayed in Fig.~\ref{fig7}. In this figure the decelerator activation time is defined with respect to $t_{\textrm{ex}}$, the center of the photoexcitation pulse applied to E1. The oscillating potentials used in recording these data were precalculated to guide atoms close to the center of the unperturbed velocity distribution, traveling at a constant speed of 1950~m/s. The oscillation frequency therefore remained constant throughout the guiding process at $\omega = 2\pi\times195$~kHz. Since only atoms that were guided away from their initial axis of propagation were detected in this experiment, the signal intensity rose when the atoms were efficiently loaded into one of the moving decelerator traps. As can be seen there are several activation times at which maxima occur in the signal associated with the guided atoms. These maxima correspond to the times at which the bunch of atoms is efficiently loaded into each consecutive moving trap of the decelerator. The temporal periodicity of these maxima corresponds to the time taken for the atoms to travel the 10~mm distance between consecutive traps, i.e., $5d_{\mathrm{cc}}/v = 5.1~\mu$s. When the decelerator is activated at the incorrect time, few atoms are guided and a minimum is observed in the integrated He$^{+}$ signal. The minima associated with these out-of-phase activation times do not reach zero intensity at times earlier than $20~\mu$s primarily because a small fraction of the $\sim10~$mm long bunch of Rydberg atoms always overlaps with one of the traps at these times when the propagation axis of the atomic beam and the axis of the decelerator coincide. At times greater than $20~\mu$s the atoms have traveled further into the decelerator when it is activated, to positions where the curved center-conductor and atomic beam axes do not overlap. In these cases, the transverse spread of the bunch of atoms is only large enough to result in guiding of a small fraction of atoms near the longitudinal positions associated with the trap minima. Away from these positions no guiding occurs and the He$^+$ signal reduces to zero. From the data in Fig.~\ref{fig7}, an activation time of $5.5~\mu$s was chosen for all experiments performed at an initial longitudinal speed $v_{0}=1950$~m/s. This permitted the full length of the decelerator to be used for guiding, acceleration and deceleration.

\begin{figure}
\begin{center}
\includegraphics[width = 7cm, angle = 0, clip=]{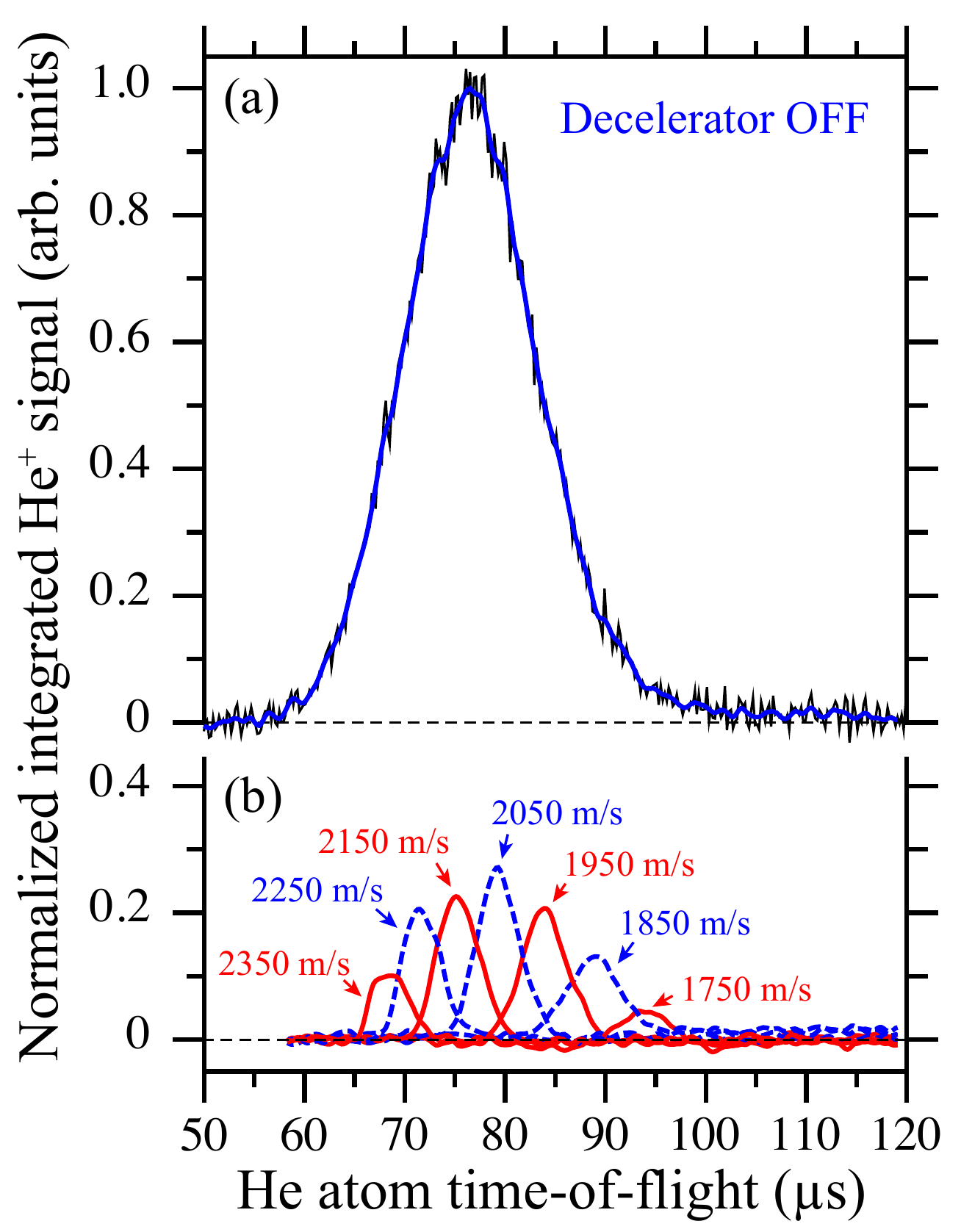}
\caption{(Color online) (a) Time-of-flight distribution of the unperturbed beam of He Rydberg atoms recorded with the decelerator off. (b) He atom time-of-flight distributions recorded with the decelerator operated to guide atoms at constant speeds ranging from 1750~m/s to 2350~m/s as indicated.}
\label{fig8}
\end{center}
\end{figure}

The ensemble of excited He Rydberg atoms prepared in the photoexcitation region has a longitudinal velocity distribution with a standard deviation of $\sim140$~m/s [see Fig.~\ref{fig8}(a)]. This corresponds to a translational temperature in the moving frame of reference of $E/k_{\textrm{B}}\sim4.7$~K. By adjusting the oscillation frequency of the decelerator potentials, isolated velocity groups within the bunch of excited atoms can therefore be guided through the device. To observe the effects of guiding at a range of speeds, measurements have been made of the time-of-flight of the He Rydberg atoms from the position of photoexcitation to the detection region, a distance of $165~$mm. The results of this set of measurements for guiding speeds from 1750~m/s to 2350~m/s, when $V_{0}=+120$~V, are displayed in Fig.~\ref{fig8}(b). In recording these data the appropriate adjustments where made to the decelerator activation time to account for the different flight-time of the atoms from their position of photoexcitation to the first trap of the decelerator at each guiding speed. From these measurements the longitudinal velocity components of the initial beam of He Rydberg atoms can be clearly identified. 

\begin{figure}
\begin{center}
\includegraphics[width = 8cm, angle = 0, clip=]{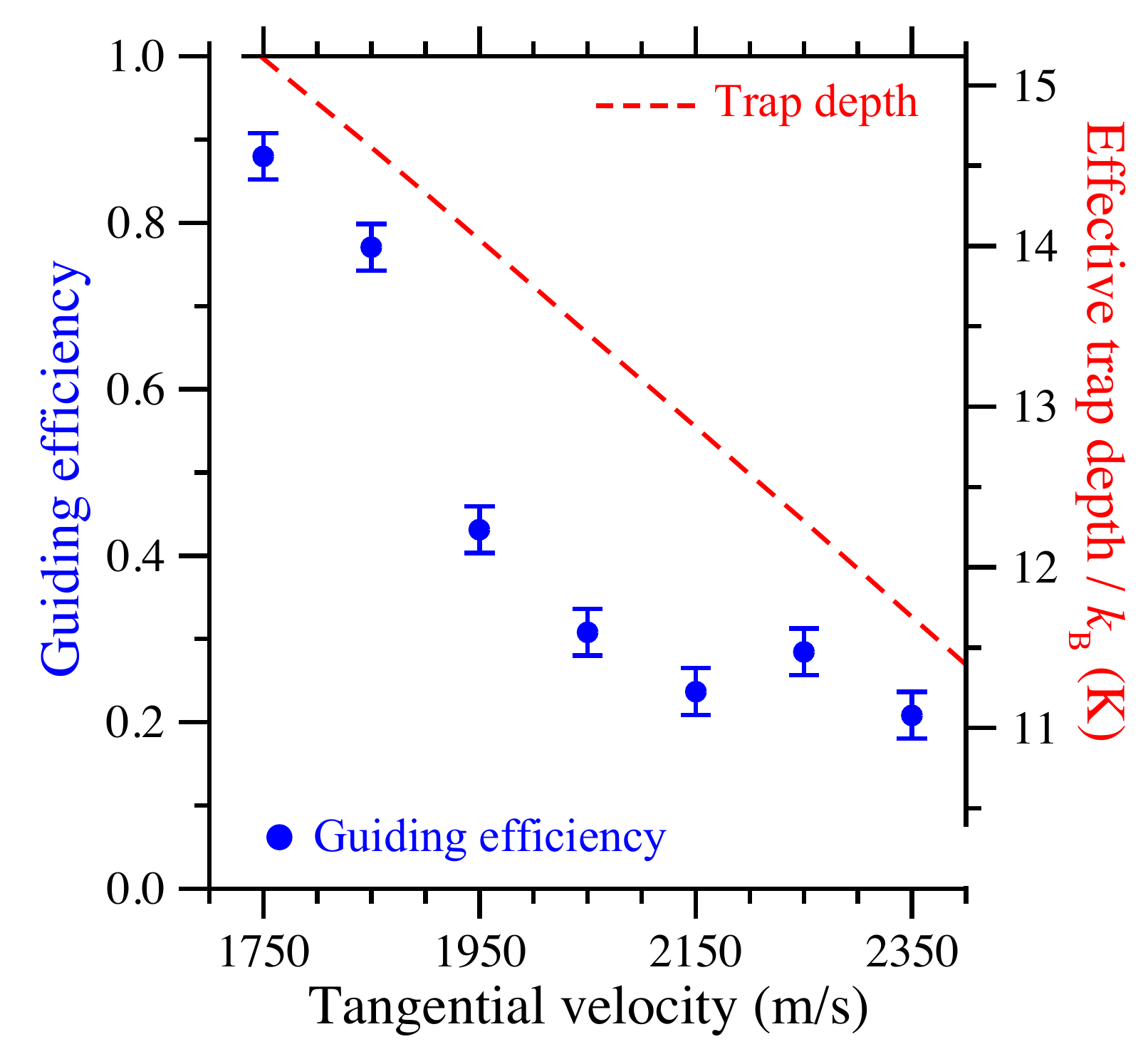}
\caption{(Color online) Dependence of the experimentally determined efficiency with which atoms were guided at constant speed through the transmission-line decelerator, upon the tangential velocity of the moving traps (blue circles - left vertical axis). The linear dependence of the effective depth of the moving traps upon their tangential velocity (see Section~\ref{sec:dec}) is indicated by the dashed red line (right vertical axis).}
\label{fig9}
\end{center}
\end{figure}

Comparison of the data in Fig.~\ref{fig8}(b) with the time-of-flight distribution of the unperturbed atomic beam, Fig.~\ref{fig8}(a), indicates that when guiding at low speeds (1750~m/s) a large fraction, $\sim0.8$, of the excited atoms are efficiently guided through the decelerator. However, close to the maximum of the unperturbed time-of-flight distribution, and for speeds approaching 2350~m/s the efficiency with which the atoms are guided is reduced to $\sim0.3$. This change in guiding efficiency can be seen more explicitly in Fig.~\ref{fig9} where individual data points, representing the ratio of the amplitude of each feature in Fig.~\ref{fig8}(b), to the amplitude of the integrated He$^+$ ion signal in Fig.~\ref{fig8}(a) at the corresponding time-of-flight, are displayed. The reduction in guiding efficiency with increasing tangential velocity is related in part to the reduction in the effective depth of the traveling electric traps in the decelerator with centripetal acceleration. However, as can be seen from the calculated linear dependence of the effective trap depth on the tangential velocity of the traps, also displayed in Fig.~\ref{fig9}, this effect alone does not completely describe the observed changes in guiding efficiency particularly for tangential velocities in the range from 1950~m/s to 2150~m/s. Because the most significant reduction in guiding efficiency is observed for these speeds close to the intensity maximum of the bunch of excited atoms, and the densities of ground state He atoms and atoms in the metastable 1s2s\,$^3\mathrm{S}_1$ state are approximately constant across the Rydberg atom cloud, the anomalies in the guiding efficiency that can be seen in Fig.~\ref{fig8} and Fig.~\ref{fig9} are thought to arise from interactions between the trapped Rydberg atoms as they travel through the decelerator. These interactions may lead to collision-induced non-adiabatic transitions to untrapped states for pairs of atoms located close to the electric field minima of the traps, as have been observed previously at lower values of $n$ in three-dimensional electrostatic traps cooled to low temperatures~\cite{seiler11a}, or to state-changing within the highly-degenerate manifold of pair states associated with the hydrogenic Rydberg-Stark states used in the experiments. Detailed investigations of these effects arising during deceleration and guiding of atoms in states with the high values of $n$ employed here, and their disentanglement from possible effects associated with the divergence of the atomic beam in free-flight, will be carried out in a future series of experiments.

\begin{figure}
\begin{center}
\includegraphics[width = 0.45\textwidth, angle = 0, clip=]{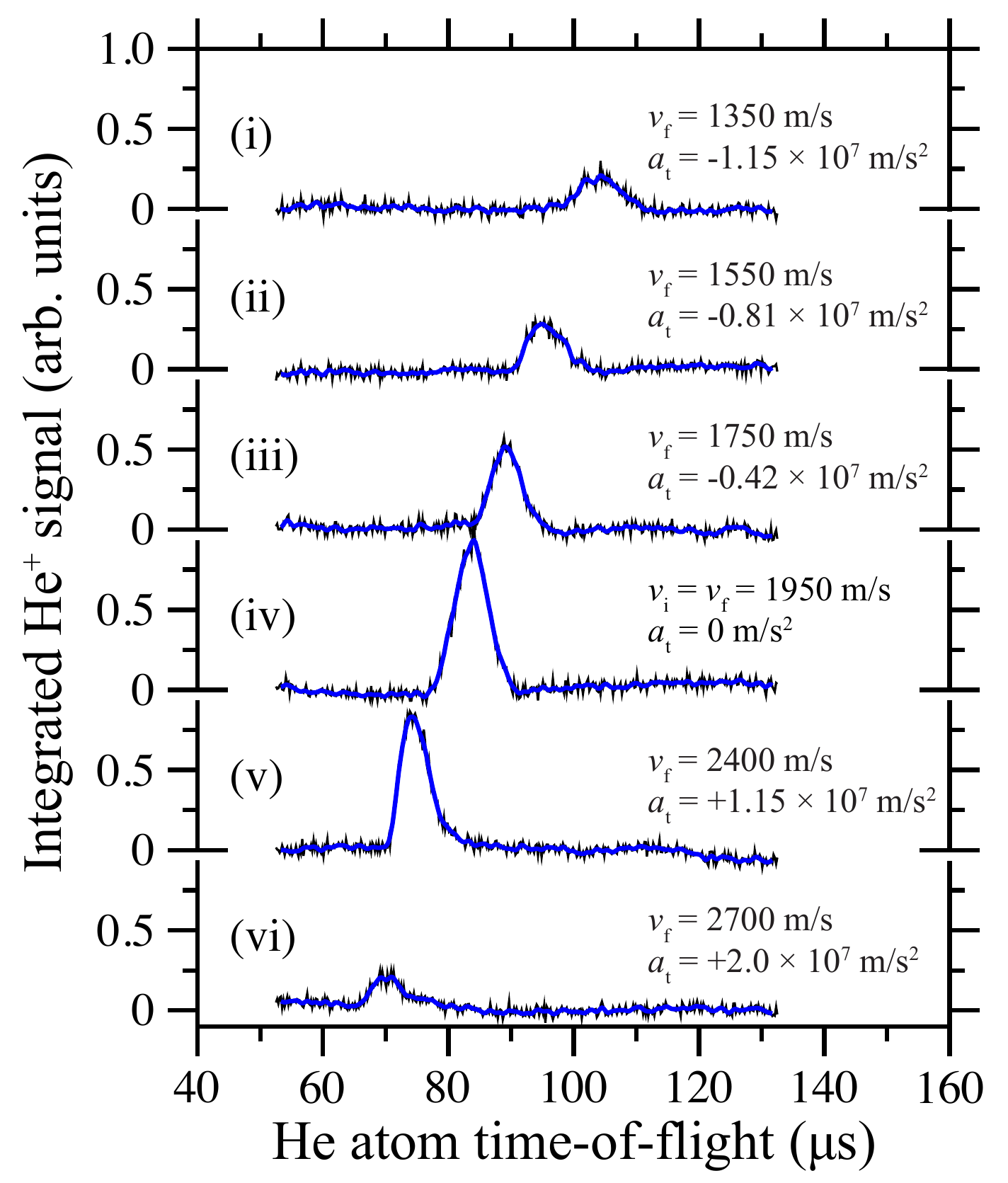}
\caption{(Color online) Acceleration and deceleration of He Rydberg atoms in the transmission-line decelerator. For reference, the time-of-flight distribution of atoms guided at a constant speed of 1950~m/s is included in dataset (iv). Accelerations as high as $a_{\textrm{t}}=+2.0\times10^7$~m/s$^2$ permit final velocities of 2700~m/s to be achieved (vi), while with $a_{\textrm{t}}=-1.15\times10^7$~m/s$^2$, deceleration to a final velocity of 1350~m/s is demonstrated (i).}
\label{fig10}
\end{center}
\end{figure}

To demonstrate acceleration and deceleration of these fast moving samples of He Rydberg atoms in the transmission-line decelerator, a component of the atomic beam with an initial longitudinal speed of 1950~m/s was loaded into a single, moving decelerator trap. The appropriate time dependence [see Eq.~(\ref{eq:timedepfreq})] was then introduced to the frequency of the oscillating decelerator potentials to exert tangential accelerations ranging from $a_{\mathrm{t}}=-1.15\times10^7$~m/s$^2$, to $a_{\mathrm{t}}=+2.0\times10^7$~m/s$^2$ on the Rydberg atoms. The results of the corresponding set of measurements are presented in Fig.~\ref{fig10}. As with the data recorded for guiding at constant speed, the time-of-flight distributions of these accelerated and decelerated samples of atoms were recorded by selectively integrating the He$^+$ signal from the atoms guided away from their initial axis of propagation as they traveled through the decelerator. In these measurements the atoms decelerated to the lowest speeds arrived at the detection region outside the time-of-flight distribution of the undecelerated beam. 

For reference, a measurement of the time-of-flight distribution of atoms guided at a constant speed of 1950~m/s is displayed in Fig.~\ref{fig10}(iv). The maximum of this distribution occurs at a flight-time of 84~$\mu$s. When a positive frequency chirp is applied to the oscillating decelerator potentials, the atoms confined in the moving traps are accelerated as they travel through the device. Applying an acceleration of $a_{\mathrm{t}}=+1.15\times10^7$~m/s$^2$, for a time of 39~$\mu$s, as the trapped atoms travel a distance of 86~mm through the decelerator leads to a final tangential velocity of 2400~m/s [Fig.~\ref{fig10}(v)]. The resulting accelerated beam arrives at the detection region 10~$\mu$s earlier than atoms guided at a constant speed of 1950~m/s. Further increasing the acceleration to $a_{\mathrm{t}}=+2.0\times10^7$~m/s$^2$, to accelerate the atoms over the same distance results in a final speed of 2700~m/s, and a further 4~$\mu$s reduction in flight-time to the detection region. At these high tangential velocities and accelerations, the effective depth of the moving decelerator traps is less than $0.1$ times the depth when guiding at a constant speed of 1950~m/s (see Fig.~\ref{fig4}). Since the bunches of Rydberg atoms fill the moving trap into which they are loaded when the decelerator is activated, the change in the amplitude of the peak of the time-of flight distribution with acceleration is, to a first approximation, expected to scale linearly with trap depth. Therefore the reduction in the amplitude of the accelerated bunch of atoms in Fig.~\ref{fig10}(vi), to $\sim0.2$ times that recorded when guiding at constant speed is close to that expected from the calculated data displayed in Fig.~\ref{fig4}.

Applying a negative frequency chirp to the decelerator waveforms leads to deceleration of the trapped atoms, and an increase in flight-time from the photoexcitation region to the detection region [Fig.~\ref{fig10}(i-iii)]. Deceleration from an initial speed of 1950~m/s over a distance of 86~mm with $a_{\mathrm{t}}=-0.42\times10^7$~m/s$^2$ ($a_{\mathrm{t}}=-0.81\times10^7$~m/s$^2$), leads to a final tangential velocity of 1750~m/s (1550~m/s) and an arrival time at the detection region $5~\mu$s ($12~\mu$s) after the bunch of atoms guided at a constant speed. Further deceleration to a final speed of 1350~m/s over the same distance, with $a_{\mathrm{t}}=-1.15\times10^7$~m/s$^2$, results in an arrival time in the detection region of $\sim104~\mu$s, $20~\mu$s later than atoms guided at 1950~m/s.

When guiding at a constant speed of 1950~m/s the effective trap depth is $E/k_{\mathrm{B}}\simeq14$~K. However, when decelerating to 1350~m/s with $a_{\mathrm{t}}=-1.15\times10^7$~m/s$^2$ the mean effective trap depth is $E/k_{\mathrm{B}}\simeq7$~K. The corresponding experimental data presented in Fig.~\ref{fig10}(iv), and Fig.~\ref{fig10}(i), exhibits a reduction to $\sim0.2$ times the intensity of the signal from atoms guided at a constant speed of 1950~m/s, upon deceleration to 1350~m/s. Similar deviations from the efficiencies expected to arise from changes in effective trap depth occur in each of the other sets of data in Fig.~\ref{fig10}. Based upon the data presented in Fig.~\ref{fig8}, the additional loss observed is thought to arise as a result of collisions between the trapped atoms during the deceleration process and a reduction in the detection efficiency of the decelerated bunches of atoms as they expand during their flight from the exit of the decelerator to the detection region.\\

\section{Discussion and Conclusion}

The transmission-line decelerator described here is a scalable device which is ideally suited to the high-speed transport, acceleration, deceleration and trapping of gas-phase samples of Rydberg atoms and molecules. Its design gives rise to symmetric electric traps which exhibit approximately equal depths in all three spatial dimensions, distinguishing it from previously developed chip-based decelerators, and permits it to be efficiently coupled to electrostatic transmission-line guides~\cite{lancuba13a}. As has been demonstrated, the array of square, surface-based electrodes that form the segmented center-conductor permit the introduction of curvatures in the plane of the surface without causing excessive distortion to the symmetry of the resulting traps. The presence of the upper plate electrode creates a very well defined electromagnetic environment within the decelerator structure. This is of particular importance for applications, including those in hybrid approaches to quantum information processing, in which inhibition of the absorption of blackbody radiation~\cite{vaidyanathan81a}, or spontaneous emission at microwave frequencies~\cite{kleppner81a}, is desirable. 

The large electric field gradients that can be generated in these transmission-line decelerators permits the controlled removal of significant amounts of kinetic energy from atomic beams over comparatively short distances and timescales. Indeed, the maximal deceleration, from 1950~m/s to 1350~m/s with $a_{\mathrm{t}}=-1.15\times10^7$~m/s$^2$, demonstrated here corresponds to a change in kinetic energy of $\Delta E_{\textrm{kin}}/e=41$~meV ($\Delta E_{\textrm{kin}}/hc=330$~cm$^{-1}$). As far as we are aware, this constitutes the largest removal of kinetic energy from a beam of Rydberg atoms or molecules reported in the literature to date.

The possibility of using transmission-line decelerators to remove large amounts of kinetic energy from samples of heavy molecules in high Rydberg states opens up exciting prospects for studies of molecular decay processes of importance in the evolution and chemical reactivity of plasmas in the upper atmosphere of the Earth~\cite{softley04a}. As can be seen in the data presented in Fig.~\ref{fig10}, the possibility of generating amplified oscillating electric potentials with frequencies approaching 1~MHz makes these transmission-line decelerators very well suited to the manipulation of atoms and molecules at speeds exceeding 2500 m/s. This will be beneficial for the rapid transport of samples within spatially separated regions of cryogenic experimental apparatus for applications in hybrid quantum information processing, and essential for experiments involving the deceleration and manipulation of beams of light positronium Rydberg atoms~\cite{cassidy14a}.

The observed effects of the density of electrically trapped Rydberg atoms on the efficiencies with which they can be manipulated when in states with the high values $n$ employed here, indicate that it will be necessary to carefully consider the role of collisions within ensembles of trapped atoms or molecules in future spectroscopy and scattering experiments where the preparation of guided or decelerated samples in selected internal quantum states is of importance. In hybrid approaches to quantum information processing involving atoms in Rydberg states with values of $n>50$, and superconducting microwave circuits, these effects may also impose a limit on the particle numbers and densities that can be employed while preserving long Rydberg-state coherence times. 
\\

\begin{acknowledgments}
This work was supported financially by the Department of Physics and Astronomy and the Faculty of Mathematical and Physical Sciences at University College London, and the Engineering and Physical Sciences Research Council under grant number EP/L019620/1.
\end{acknowledgments}

\end{document}